\documentclass{llncs}
\usepackage{amsfonts}

\begin{document}

\title{On the brightness of the Thomson lamp.\\ A prolegomenon to quantum recursion theory}
\titlerunning{On the brightness of the Thomson lamp. A prolegomenon to quantum recursion theory}

\author{Karl Svozil}
\institute{Institute for Theoretical Physics, Vienna University of Technology,  \\
Wiedner Hauptstra\ss e 8-10/136, A-1040 Vienna, Austria \\
{http://tph.tuwien.ac.at/\homedir svozil}\\
\email{svozil@tuwien.ac.at}
}


\maketitle

\begin{abstract}
Some physical aspects related to the limit operations of the Thomson lamp are discussed. Regardless of the formally unbounded and even infinite number of ``steps'' involved, the physical limit has an operational meaning in agreement with the Abel sums of infinite series. The formal analogies to accelerated (hyper-) computers and the recursion theoretic diagonal methods are discussed. As quantum information is not bound by the mutually exclusive states of classical bits, it allows a consistent representation of fixed point states of the diagonal operator. In an effort to reconstruct the self-contradictory feature of diagonalization, a generalized diagonal method allowing no quantum fixed points is proposed.
\end{abstract}

\section{Introduction}

{\it Caveats} at the very beginning of a manuscript may appear overly prudent and displaced,
stimulating even more {\it caveats} or outright rejection.
Yet, one should keep in mind that, to rephrase a {\it dictum} of John von Neumann~\cite{von-neumann1},
from an operational point of view~\cite{bridgman},
anyone who considers physical methods of producing infinity is, of course, in a state of sin.
Such sinful physical pursuits qualify for Neils Henrik Abel's verdict that
(Letter to Holmboe, January 16, 1826~\cite{Hardy:1949,rousseau-2004}),
{\em ``divergent series are the invention of the devil, and it is shameful to base on them any demonstration whatsoever.''}
This, of course, has prevented no-one, in particular not Abel himself, from considering these issues.
Indeed, by reviving old eleatic ideas~\cite{zeno,salmon-01},
accelerated computations~\cite{weyl:49} have been the main paradigm of
the fast growing field of hypercomputations
(e.g., Refs.~\cite{Davis-2006,Doria-2006,ord-2006}).
For the sake of hypercomputation,
observers have been regarded on
their path toward black holes~\cite{Hogarth92,DBLP:conf/mcu/Durand-Lose04,Nemeti2006118},
and automata have been densely embedded~\cite{schaller-09}, to name just two ``mind-boggling'' proposals.

In what follows we shall discuss issues related to the quasi-physical aspects of hypercomputers,
or more specifically, accelerated agents or processes approaching the limit of infinite computation.
Thereby, we shall not be concerned with issues related to unbounded space or memory
consumption discussed by Calude and Staiger~\cite{calude-staiger-09},
although we acknowledge their importance.
In analogy to Benacerraf's discussion~\cite{benna:62} of Thomson's proposal~\cite{thom:54,thom:54a}
of a lamp which is switched on or off at
geometrically decreasing time delays, Shagrir~\cite{1011191} suggested that the physical state has little or no relation to
the states in the previous acceleration process.

This argument is not dissimilar to Bridgman's~\cite{bridgman} argument against the use of Cantor-type diagonalization
procedures on the basis that it is physically impossible to operationalize and construct some process ``on top of'' or after
a non-terminating, infinite process~\cite{Hamkins-02}.
The method of diagonalization presents an important technique of recursion theory~\cite{rogers1,odi:89,Boolos-07}.
Some aspects of its physical realizability have already been discussed by the author~\cite{maryland,quantum-omega,svo-1995-paradox,sv-aut-rev}.
In what follows we shall investigate some further physical issues related to ``accelerated'' agents and
the operationalizability of the diagonalization operations in general.

\section{Classical brightness of the Thomson lamp}

The Thomson lamp is some light source which is switched off when it is on, and conversely switched on when it is off;
whereby the switching intervals are geometrically (by a constant factor smaller than unity) reduced or ``squeezed.''
By construction, the switching process never ``saturates,'' since the state always flips from ``on'' to ``off''
and then back to ``on'' again {\em ad infinitum;}
yet, as this infinity of switching cycles is supposed to be reachable in finite time,
one may wonder what happens at and after the accumulation point.

The Thomson process can formally be described at {\em intrinsic} (unsqueezed, unaccelerated)
discrete time steps $t$; i.e., by the partial sum
\begin{equation}
s(t)=\sum_{n=0}^t (-1)^n = \left\{
\begin{array}{ll}
1 &\textrm{for even $t$,}
\\ 0 &\textrm{for odd $t$,}
\end{array}
\right.
\end{equation}
which can be interpreted as the result of all switching operations until time $t$.
The intrinsic time scale is related to an {\em extrinsic} (squeezed, accelerated) time scale~\cite[p.~26]{svozil-93}
\begin{equation}
 \tau_0=0,\;
 \tau_{t>0}=\sum_{n=1}^t 2^{-n}=2\left(1-2^{-t}\right)\quad .
\end{equation}
In the limit
 of infinite intrinsic time $t\rightarrow \infty$, the proper time
 $\tau_\infty = 2$ remains finite.

If one encodes the physical states of the Thomson lamp by ``0'' and ``1,''
associated with the lamp ``on'' and ``off,'' respectively, and the switching process with the concatenation of ``+1'' and ``-1'' performed so far, then
the divergent infinite series associated with the Thomson lamp is
the Leibniz series~\cite{leibnitz-1860,moore-1938,Hardy:1949,everest-2003}
\begin{equation}
s = \sum_{n=0}^\infty (-1)^n=1-1+1-1+1-\cdots \stackrel{{\rm A}}{=} \frac{1}{2}.
\label{2009-fiftyfifty-1}
\end{equation}
Here, ``A'' indicates the Abel sum~\cite{Hardy:1949} obtained from a ``continuation'' of the geometric series, or alternatively, by
$s= 1-s$.
As this shows, formal summations of the Leibnitz type (\ref{2009-fiftyfifty-1}) require specifications
which could make them unique.

Moreover, divergent sums could be valuable {\em approximations} of solutions of differential equations
which might be even {\em analytically solvable}.
One of the best-known examples of this case is Euler's differential equation~\cite{balser-2000,rousseau-2004,Costin-2009}
$z^2y'+y=z$, which has both (i) a formal solution as a power series ${\hat{f}}(z)=-\sum_{n=0}^\infty n!\,(-z)^n$,
which {\em diverges} for all nonzero values of $z$;
as well as (ii) an exact solution
${\hat{f}}(z) =e^{1/z}{\rm Ei}(-1/z)$ obtained by direct integration.
${\rm Ei}$ stands for the exponential integral ${\rm Ei}(z)=\int_{z}^\infty e^{-t}/t$.
The difference between the exact and the series solution up to index $k$ can be approximated~\cite{rousseau-2004} by
$\vert  f(z)-f_k(z)\vert \le \sqrt{2\pi z}e^{-1/z}$, which is exponentially small for small $z$.
To note another example, Laplace has successfully (with respect to the predictions made) used divergent
power series for calculations of planetary motion.
Thus it is not totally unreasonable to suspect that the perturbation series solutions obtained via diagrammatical
techniques~\cite{Itzykson-zuber,lurie,thooft-Veltman_Diagrammar}
for the differential equations of quantum electrodynamics are of this ``well behaved'' type.

Therefore,  at least in principle, the Thomson lamp could be perceived as a physical process governed by some associated differential equation,
such as $y'(1-z)-1=0$, which has an exact solution $f(z)=\log (1/1-x)$; as well as a divergent series solution  ${\hat{f}}(z)=\sum_{n=0}^\infty (-z)^n/n$,
so that the first derivative of ${\hat{f}}(z)$, taken at $z=1$, yields the Leibniz series $s$.

If one is willing to follow the formal summation to its limit,
then the Abel sum could be justified from the point of view of classical physics as follows.
Suppose that the switching processes can be disregarded,
an assumption not dissimilar to related concerns for Maxwell's demon~\cite{maxwell-demon}.
Suppose further that all measurements are finite in the sense that the temporal resolution $\delta$ of the observation of the Thomson lamp cannot be made ``infinitely small;''
i.e., the observation time is finitely limited from below by some arbitrary but non-vanishing time interval.
Then, the mean  brightness of the Thomson lamp can be operationalized by measuring the {\em average} time of the lamp to be ``on'' as compared to
being ``off'' in the time interval determined by the  temporal resolution of the measuring device.
This definition is equivalent to an idealized ``photographic'' plate or light-sensitive detector which additively collects light during the time the shutter is open.
Then, such a device would record the cycles of Thomson's lamp up to temporal resolution $\delta$, and would then integrate the remaining light emanating from it.
Suppose that $\delta = 1/2^t$, and that at the time $\sum_{n=1}^t 2^{-n}=2\left(1-2^{-t}\right)$, the Thomson lamp is in the ``off'' state.
Then the ``on'' and ``off'' periods sum up to
\begin{equation}
\begin{array}{lclllllllllllllllllll}
s_0&=&  &    \frac{1}{4}   &+&    \frac{1}{16} &+&    \frac{1}{64}  &+& \cdots  &=&   \frac{1}{3}   \\
s_1&=&\frac{1}{2}   &+&    \frac{1}{8} &+&    \frac{1}{32}     &+& \cdots &\cdots  &=&   \frac{2}{3} \; .
\end{array}
\end{equation}
By averaging over the initial time, there is a 50:50 chance that the Thomson lamp is ``on'' or ``off'' at highest resolution --- which is equivalent to
repeating the experiment with different offsets -- resulting in an average brightness of $1/2$, which is identical to the Abel sum.
In this way, the Abel sum can be justified from a physical point of view.
The use of the Abel sum may not be perceived as totally convincing, as the Abel sum can be affected by changes to an initial segment
of the series.

In a strict sense, this classical physics treatment of the brightness of the Thomson lamp is of little use for predictions of the state of the Thomson lamp {\em after} the limit of the switching cycles.
One might still put forward that, in accordance with our operationalization of the average brightness, there is a 50:50 probability that it is in state ``on'' or ``off.''
Just as for the Thomson lamp discussed above, one needs to be careful in defining the output of
an accelerated Turing
machine~\cite{svozil-93,maryland,quantum-omega,svo-1995-paradox,sv-aut-rev,Copeland98eventuring,potgieter-06,1011191}.

\section{Quantum state of the Thomson lamp}

The quantum mechanical formalization of the Thomson lamp presents an intriguing possibility: as quantum information is not bound by two classically contradictory states,
the quantum state of the Thomson lamp can be in a superposition thereof.
At first glance, the fact that quantum information co-represents contradictory cases appears ``mind-boggling'' at best.
As Hilbert pointedly stated (in Ref.~\cite[p.~163]{hilbert-26}, he was not discussing quantum information theory):
{\em ``$\ldots$~the conception that facts and events could contradict
themselves appears to me as an exemplar of thoughtlessness\footnote{
{\em ``$\ldots$~mir erscheint die Auffassung, als k{\"{o}}nnten Tatsachen und Ereignisse selbst miteinander in Widerspruch
geraten, als das Musterbeispiel einer Gedankenlosigkeit.''}}.''}
Nevertheless,  ``quantum parallelism'' appears to be an important property of quantum information and computation,
which is formally expressed by a coherent summation of orthogonal pure states.
This property can be used for quantum states which are formally invariant ``fixed points'' of diagonalization operators.

We note a particular
aspect of the quantization of the electromagnetic field emanating from the Thomson lamp:
the quanta recorded at the photographic emulsion or detector
impinge on these light sensitive devices not continuously but in discrete ``energy lumps'' of $E_\nu = h \nu = hc/\lambda$,
where $\nu$ and $\lambda$ refer to the frequency
and wavelength of the associated field mode.
We shall also not be concerned with the photon statistics,
which would require a detailed analysis of the light source~\cite{Mandel-Wolf}.

In what follows, the notation of Mermin~\cite{mermin-07} will be used.
Let $\vert 0\rangle = (1,0)$ and $\vert 1\rangle = (0,1)$ be the representations of the ``off'' and ``on''
states of the Thomson lamp, respectively.
Then, the switching process can be symbolized by the ``not'' operator
$\textsf{\textbf{X}}=
\left(
\begin{array}{cc}
0&1\\
1&0
\end{array}
\right)$, transforming $\vert 0\rangle$ into $\vert 1\rangle$ and {\it vice versa}.
Thus the quantum switching process of the Thomson lamp at time $t$ is the partial product~\cite[Sect.~5.11]{arfken05}
\begin{equation}
\textsf{\textbf{S}}(t)
= \prod_{n=0}^t
\textsf{\textbf{X}}^t=   \left\{
\begin{array}{ll}
{\Bbb I}_2 &\textrm{for even $t$,}
\\ \textsf{\textbf{X}} &\textrm{for odd $t$.}
\end{array}
\right.
\end{equation}
In the limit, one obtains an infinite product $\textsf{\textbf{S}}$ of matrices with the two accumulation points mentioned above.

The eigensystem of $\textsf{\textbf{S}}(t)$ is given by the two 50:50 mixtures of $\vert 0\rangle $ and $\vert 1\rangle $
with the two eigenvalues $1$ and $-1$; i.e.,
\begin{equation}
\textsf{\textbf{S}}(t)\frac{1}{\sqrt{2}}\left( \vert 0\rangle \pm \vert 1\rangle   \right)
= \pm \frac{1}{\sqrt{2}}\left( \vert 0\rangle \pm \vert 1\rangle   \right) = \pm \vert \psi_\pm \rangle.
\end{equation}
In particular,  the state   $\vert \psi_+ \rangle$ associated with the eigenvalue $+1$
is a {\em fixed point} of the operator $\textsf{\textbf{S}}(t)$.
These are the two states which can be expected to emerge as the quantum state of the Thomson lamp
in the limit of infinity switching processes.
Note that, as for the classical case and for the formal Abel sum, they represent a fifty-fifty mixture of the ``on'' and ``off'' states.

\section{Quantum fixed point of diagonalization operator}

In set theory, logic and in recursion theory~\cite{rogers1,davis,Barwise-handbook-logic,enderton72,odi:89,Boolos-07},
the method of {\em proof by contradiction} ({\it reductio ad absurdum}) requires some  ``switch of the bit value,'' which is applied in a self-reflexive manner.
Classically, total contradiction is achieved by supposing that some proposition is true,
then proving that under this assumption it is false, and {\it vice versa}.
The general proof method thus can be intuitively related to the application of a ``not'' operation.

The following considerations rely on the reader's informal idea of effective computability
and algorithm, but
they can be translated into a rigorous description involving universal computers such as
Turing machines~\cite[Chapter~C.1]{Barwise-handbook-logic}.
Assume that an algorithm $A$ is
restricted to classical bits of information.
For the sake of contradiction,
assume that there exists a (suspicious) halting algorithm $ h $ which outputs the code of a classical bit as follows:
\begin{equation}
 h  ( B(X) ) =     \left\{
\begin{array}{ll}
0 &\textrm{ whenever $B(X)$ does not halt},  \\
1 &\textrm{  whenever $B(X)$ halts}.
\end{array}
\right.
\label{el:halt}
\end{equation}

Then, suppose an algorithm $A$ using $ h $ as a subprogram,
which performs the diagonal argument
by halting whenever (case 1) $ h (B(X))$ indicates that $B(X)$ does not halt (diverges),
and conversely (case 2)
by not halting whenever $ h (B(X))$ indicates that $B(X)$ halts (converges).
Self-reflexivity reveals the antinomy of this construction:
upon operating on its own code, $A$ reaches at a total contradiction:
whenever  $A(A)$
halts, $ h (A(A))$ outputs $1$ and forces $A(A)$ not to halt.
Conversely,
whenever $A(A)$ does not halt, then $ h (A(A))$ outputs $0$
and steers
$A(A)$ into the halting mode.  In both cases one arrives at a complete
contradiction.  Classically, this contradiction can only be consistently
avoided by assuming the non-existence of $A$ and, since the only
non-trivial feature of $A$ is the use of the peculiar halting algorithm
$ h $, the impossibility of any such halting algorithm.

As has already been argued~\cite{maryland,quantum-omega,svo-1995-paradox,sv-aut-rev},
$\vert \psi_+ \rangle = ({1}/{\sqrt{2}})\left( \vert 0\rangle + \vert 1\rangle   \right)$
is the quantum fixed point state of the ``not'' operator,
which is essential for diagonal arguments, as
$
\textsf{\textbf{X}}\vert \psi_+ \rangle = \vert \psi_+ \rangle
$.
Thus in quantum recursion theory,
the diagonal argument consistently goes through without leading to a contradiction,
as $ h (A(A))$ yielding $\vert \psi_+ \rangle$ still allows a consistent response
of $A$ by a coherent superposition of its halting and non-halting states.
It should be noted, however, that the fixed point quantum ``solution''
of the  halting problem cannot be utilized.
In particular, if one is interested in the ``classical'' solution of the decision problem whether
or not $A(A)$ halts,  then one ultimately has to perform an
irreversible measurement
on the fixed point state. This  causes a state reduction into the
classical states corresponding to $\vert 0\rangle$ and $\vert 1\rangle$.
Any single measurement  yields an indeterministic result:
According to the Born rule
(cf. \cite[p.~804]{born-26-2}, English translation in \cite[p.~302]{jammer:89} and~\cite{zeil-05_nature_ofQuantum}),
when measured ``along'' the classical basis (observable) $\{\vert 0\rangle,\vert 1\rangle \}$,
the probability that the fixed point state $\vert \psi_+ \rangle$  returns at random
one of the two classical states $\vert 0\rangle$ (exclusive) or $\vert 1\rangle$, is $1/2$.
Thereby, classical undecidability is recovered.
Thus, as far as problem solving is concerned, this method involving quantum information does not present an advantage
over classical information.  For the general case discussed, with regards to the question of
whether or not a computer halts, the quantum ``solution'' fixed point state
is equivalent to the throwing of a fair classical coin~\cite{diaconis:211}.

\section{Quantum diagonalization}

The above argument used the continuity of quantum bit states as compared to the discrete classical spectrum of just
two classical bit states for a construction of fixed points of the
diagonalization operator modeled by $\textsf{\textbf{X}}$. One could proceed a step further and allow
{\em non-classical diagonalization procedures}. Thereby, one could allow
the entire range of two-dimensional unitary transformations~\cite{murnaghan}
\begin{equation}
\textsf{\textbf{U}}_2(\omega ,\alpha ,\beta ,\varphi )=e^{-i\,\beta}\,
\left(
\begin{array}{cc}
{e^{i\,\alpha }}\,\cos \omega
&
{-e^{-i\,\varphi }}\,\sin \omega
\\
{e^{i\,\varphi }}\,\sin \omega
&
{e^{-i\,\alpha }}\,\cos \omega
 \end{array}
\right)
 \quad ,
\label{e:quid3}
\end{equation}
where $-\pi \le \beta ,\omega \le \pi$,
$-\, {\pi \over 2} \le  \alpha ,\varphi \le {\pi \over 2}$, to act on
the quantum bit.
A typical example of a non-classical operation on a quantum bit is
the ``square root of not''
($
\sqrt{\textsf{\textbf{X}}}
\cdot
\sqrt{\textsf{\textbf{X}}} =\textsf{\textbf{X}}$) gate operator
\begin{equation}
\sqrt{\textsf{\textbf{X}}} =
{1 \over 2}
\left(
\begin{array}{cc}
1+i&1-i
\\
1-i&1+i
 \end{array}
\right)
\quad ,
\end{equation}
which again has the fixed point state $\vert \psi_+ \rangle$ associated with the eigenvalue $+1$.
Yet, not all of these unitary transformations have eigenvectors
associated with eigenvalues $+1$ and thus fixed points.
Indeed, only
unitary transformations of the form
\begin{equation}
\begin{array}{c}
[\textsf{\textbf{U}}_2(\omega ,\alpha ,\beta ,\varphi )]^{-1}\,\mbox{diag}(1, e^{i\lambda
}) \textsf{\textbf{U}}_2(\omega ,\alpha ,\beta ,\varphi )=
\qquad
\qquad
\qquad
\qquad
\qquad  \\
=
\left(
\begin{array}{cc}
\cos^2 \omega  + e^{i\,\lambda} \,\sin^2 \omega
&
{1\over 2}
e^{-i\,\left(\alpha +\varphi \right) }
(e^{i\,\lambda}-1)
\sin (2\,\omega )
\\
{1\over 2}
 e^{i\,\left(\alpha +\varphi \right)}
(e^{i\,\lambda }-1)
\sin (2\,\omega )
&
e^{i\,\lambda }\,\cos^2 \omega  + \sin^2 \omega
 \end{array}
\right)
\end{array}
\end{equation}
for arbitrary $\lambda$ have fixed points.

Applying non-classical operations on quantum bits with no fixed points
\begin{equation}
\begin{array}{c}
[\textsf{\textbf{U}}_2(\omega ,\alpha ,\beta ,\varphi )]^{-1}\,\mbox{diag}( e^{i\mu } ,
e^{i\lambda }) \textsf{\textbf{U}}_2(\omega ,\alpha ,\beta ,\varphi ) =
\qquad
\qquad
\qquad
\qquad
\qquad  \\
=
\left(
\begin{array}{cc}
e^{i\,\mu }\,\cos^2 \omega  + e^{i\,\lambda }\,\sin^2 \omega
&
{1\over 2}
e^{-i\,\left( \alpha  + \varphi \right) }
\left( e^{i\,\lambda } - e^{i\,\mu } \right)
\sin (2\,\omega )
\\
{1\over 2}
e^{i\,\left( \alpha  + \varphi \right)}
\left( e^{i\,\lambda } - e^{i\,\mu }  \right) \,\sin (2\,\omega )
&
e^{i\,\lambda } \cos^2 \omega  + e^{i\,\mu }\sin^2 \omega
 \end{array}
\right)
\end{array}
\end{equation}
with $\mu ,\lambda \neq 2n\pi$, $n\in {\Bbb N}_0$ gives rise to
eigenvectors which are not fixed points, and which acquire non-vanishing
phases $\mu , \lambda$ in the generalized diagonalization process.

\begin{figure}
\begin{center}
\unitlength .5mm 
\linethickness{0.4pt}
\ifx\plotpoint\undefined\newsavebox{\plotpoint}\fi 
\begin{picture}(120,80)(0,0)
\put(20,0){\framebox(80,80)[cc]{}}
\put(57.67,40){\line(1,0){5}}
\put(64.33,40){\line(1,0){5}}
\put(50.67,40){\line(1,0){5}}
\put(78.67,50){\dashbox{1,1}(8,4.33)[cc]{}}
\put(82.67,58){\makebox(0,0)[cc]{$P_3,\varphi$}}
\put(73.33,40){\makebox(0,0)[lc]{$S(T(\omega ))$}}
\put(8.33,65.67){\makebox(0,0)[cc]{$\vert 0\rangle$}}
\put(110.67,65.67){\makebox(0,0)[cc]{${\vert 0\rangle}'$}}
\put(110.67,25.67){\makebox(0,0)[cc]{${\vert 1\rangle}'$}}
\put(8,25.67){\makebox(0,0)[cc]{$\vert 1\rangle$}}
\put(24.33,75.67){\makebox(0,0)[lc]
{${\textsf{\textbf{U}}}^{bs}(\omega ,\alpha ,\beta ,\varphi )$}}
\put(0,59.67){\vector(1,0){20}}
\put(0,20){\vector(1,0){20}}
\put(100,60){\vector(1,0){20}}
\put(100,20){\vector(1,0){20}}
\put(20,20){\line(2,1){80}}
\put(20,60){\line(2,-1){80}}
\put(32.67,50){\dashbox{1,1}(8,4.33)[cc]{}}
\put(36.67,62){\makebox(0,0)[cc]{$P_1,\alpha +\beta $}}
\put(32.67,27){\dashbox{1,1}(8,4.33)[cc]{}}
\put(36.67,35){\makebox(0,0)[cc]{$P_2,\beta$}}
\end{picture}
\end{center}
\caption{Quantum gate operating on a qubit
realized by a four-port interferometer with two input ports ${\vert 0\rangle} ,{\vert 1\rangle} $,
a beam splitter $S(T)$ with variable transmission $T(\omega )$,
three phase shifters $P_1,P_2,P_3$,
and two output ports ${\vert 0\rangle} ',{\vert 1\rangle} '$.
 \label{2009-fiftyfifty}}
\end{figure}
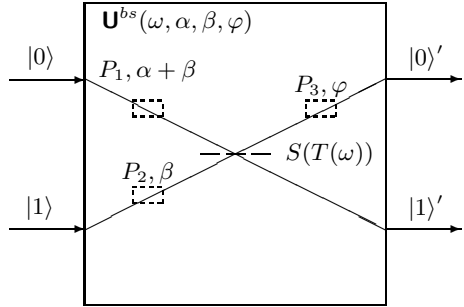
For the sake of demonstration of a physical realization,  consider an elementary diagonalization operator without a fixed point and with equal phases
in the diagonal terms; i.e., $\mu = \lambda$,
thus reducing to $\mbox{diag}( e^{i\lambda } , e^{i\lambda })$.
The generalized quantum optical beam splitter sketched in Fig.~\ref{2009-fiftyfifty}
can be described either by the transitions~\cite{green-horn-zei}
\begin{equation}
\begin{array}{rlcl}
P_1:&\vert {0}\rangle  &\rightarrow& \vert {0}\rangle e^{i(\alpha +\beta)}
 , \\
P_2:&\vert {1}\rangle  &\rightarrow& \vert {1}\rangle
e^{i \beta}
, \\
S:&\vert {0} \rangle
&\rightarrow& \sqrt{T}\,\vert {1}'\rangle  +i\sqrt{R}\,\vert {0}'\rangle
, \\
S:&\vert {1}\rangle  &\rightarrow& \sqrt{T}\,\vert {0}'\rangle  +i\sqrt{R}\,\vert
{1}'\rangle
, \\
P_3:&\vert {0}'\rangle  &\rightarrow& \vert {0}'\rangle e^{i
\varphi
} ,
\end{array}
\end{equation}
where every reflection by a beam splitter $S$ contributes a phase $\pi /2$
and thus a factor of $e^{i\pi /2}=i$ to the state evolution.
Transmitted beams remain unchanged; i.e., there are no phase changes.

Alternatively, with
$\sqrt{T(\omega )}=\cos \omega$
and
$\sqrt{R(\omega )}=\sin \omega$,
the action of a lossless beam splitter may be
described by the matrix\footnote{
The standard labeling of the input and output ports are interchanged.
Therefore, sine and cosine functions are exchanged in the transition matrix.}
$$
\left(
\begin{array}{cc}
i \, \sqrt{R(\omega )}& \sqrt{T(\omega )}
\\
\sqrt{T(\omega )}&  i\, \sqrt{R(\omega )}
 \end{array}
\right)
=
\left(
\begin{array}{cc}
i \, \sin \omega  & \cos \omega
\\
\cos \omega&  i\, \sin \omega
 \end{array}
\right)
.
$$
A phase shifter is represented by
either
$\mbox{diag}\left(
e^{i\varphi },1
\right)
$
or
$\mbox{diag}
\left(
1,e^{i\varphi }
\right)
$ in two-dimensional Hilbert space.
 The action of the entire device consisting of such elements is
calculated by multiplying the matrices in reverse order in which the
quanta pass these elements \cite{yurke-86,teich:90}; i.e.,
\begin{equation}
\begin{array}{l}
{\textsf{\textbf{U}}}^{bs} (\omega ,\alpha ,\beta ,\varphi )=
\left(
\begin{array}{cc}
e^{i\varphi}& 0\\
0& 1
\end{array}
\right)
\left(
\begin{array}{cc}
i \, \sin \omega  & \cos \omega
\\
\cos \omega&  i\, \sin \omega
\end{array}
\right)
\left(
\begin{array}{cc}
e^{i(\alpha + \beta)}& 0\\
0& 1
\end{array}
\right)
\left(
\begin{array}{cc}
1&0\\
0& e^{i\beta }
\end{array}
\right)
\\
\qquad
\qquad
\qquad
\qquad
\qquad
=
\left(
\begin{array}{cc}
 i \,e^{i \,\left( \alpha + \beta + \varphi \right) }\,\sin \omega &
   e^{i \,\left( \beta + \varphi \right) }\,\cos \omega
\cr
   e^{i \,\left( \alpha + \beta \right) }\, \cos \omega&
i \,e^{i \,\beta}\,\sin \omega
 \end{array}
\right)
.
\end{array}
\end{equation}
For this physical setup, the phases
$\omega =\pi /2$,
$\beta = \lambda -\pi /2$
and $\varphi =-\alpha$
can be arranged such that
${\textsf{\textbf{U}}}^{bs} (\pi /2,\alpha , \lambda -\pi /2 ,-\alpha )= \mbox{diag}( e^{i\lambda } , e^{i\lambda })$.
Another example is
${\textsf{\textbf{U}}}^{bs} (\pi /2,2\lambda , -\pi /2 -\lambda ,0)= \mbox{diag}( e^{i\lambda } , e^{-i\lambda })$.
For the physical realization of general unitary operators in terms of beam splitters
the reader is referred to Refs.~\cite{rzbb,reck-94,zukowski-97,svozil-2004-analog}.

\section{Summary}

In summary we have discussed some physical aspects related to the limit operations of the Thomson lamp.
This physical limit, regardless of the formally unbounded and even infinite number of ``steps'' involved,
has an operational meaning in agreement with the formal Abel sums of infinite series.
We have also observed the formal analogies to accelerated (hyper-)computers and have discussed the recursion theoretic diagonal methods.
As quantum information is not bound by mutually exclusive states of classical bits,
it allows a consistent representation of fixed point states of the diagonal operator.
In an effort to reconstruct the self-contradictory feature of diagonalization and the resulting {\it reductio ad absurdum},
a generalized diagonal method allowing no quantum fixed points has been proposed.

$\;$\\
{\bf Acknowledgements}
\\
The author gratefully acknowledges discussions with Cristian Calude,
Liam Fearnley andMartin Ziegler, as well as the kind hospitality of the {\it Centre for Discrete Mathematics
and Theoretical Computer Science (CDMTCS)} of the {\it Department of Computer Science at
The University of Auckland.}


\end{document}